\documentclass{article}

\usepackage{arxiv}

\usepackage[utf8]{inputenc} % allow utf-8 input
\usepackage[T1]{fontenc}    % use 8-bit T1 fonts
\usepackage{hyperref}       % hyperlinks
\usepackage{url}            % simple URL typesetting
\usepackage{booktabs}       % professional-quality tables
\usepackage{amsfonts}       % blackboard math symbols
\usepackage{nicefrac}       % compact symbols for 1/2, etc.
\usepackage{microtype}      % microtypography
\usepackage{lipsum}		% Can be removed after putting your text content
\usepackage{graphicx}
\usepackage{doi}
\usepackage{graphicx}% Include figure files
\usepackage{dcolumn}% Align table columns on decimal point
\usepackage{bm}% bold math
\usepackage{amssymb}
\usepackage{amsthm}
\usepackage{amsmath}
\DeclareMathOperator{\Tr}{Tr}
\newtheorem{theorem}{Theorem}

\newtheorem{proposition}[theorem]{Proposition}
\newtheorem{observation}{Observation}

\usepackage{enumerate}

\title{Filtering of higher-dimensional entanglement networks using information volumes}

\author{ \href{https://orcid.org/0000-0001-5772-3937}{\includegraphics[scale=0.06]{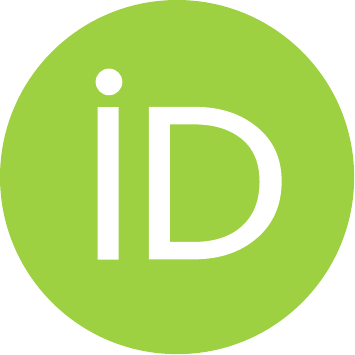}\hspace{1mm}Shahabeddin M.~Aslmarand} \\
	Department of Physics\\
	Florida Atlantic University\\
	 Boca Raton, FL 33431, USA\\
	
	\texttt{} \\
	\And
	\href{https://orcid.org/0000-0002-5883-3596}{\includegraphics[scale=0.06]{orcid.pdf}\hspace{1mm}Warner A.~Miller} \\
		Department of Physics\\
	Florida Atlantic University\\
	 Boca Raton, FL 33431, USA\\
	\texttt{} \\
		\And
	\href{}{\includegraphics[scale=0.06]{orcid.pdf}\hspace{1mm}Doyeol (David) Ahn} \\
	Center Quantum Information Processing\\ Department of Electrical and Computer Engineering\\ University of Seoul,\\ ${}^{}$ Seoul 130-743, Republic of Korea.
	
		\And
	\href{}{\includegraphics[scale=0.06]{orcid.pdf}\hspace{1mm}Paul M. Alsing} \\
	Air Force Research Laboratory\\ Information Directorate\\ Rome, NY 13441, United States of America.
	\texttt{} \\
}

\renewcommand{\shorttitle}{A geometric representation of entanglement}

\hypersetup{
pdftitle={A geometric representation of entanglement},
pdfauthor={Shahabeddin M.~Aslmarand, Warner A.~Miller},
pdfkeywords={Quantum Entanglement, Quantum Correlation, Quantum mechanics},
}

\begin{document}
\maketitle

\begin{abstract}
We introduce a novel geometric approach to characterize entanglement relations in large quantum systems. Our approach is inspired by Schumacher's singlet state triangle inequality, which used an entropic-based distance to capture the strange properties of  entanglement using geometric-based inequalities.  Schumacher uses classical entropy and can only describe the geometry of bipartite states.  We extend his approach by using von Neumann entropy to create an entanglement monotone that can be generalized for higher dimensional systems. We achieve this by utilizing recent definitions for entropic areas, volumes, and higher dimensional volumes for multipartite quantum systems. This enables us to differentiate systems with high quantum correlation from systems with low quantum correlation and differentiate between different types of multi-partite entanglement.  It also enable us to describe some of the strange properties of quantum entanglement using simple geometrical inequalities. Our geometrization of entanglement provides  new  insight into quantum entanglement. Perhaps by constructing well motivated geometrical structures (e.g. relations among areas, volumes ...), a set of trivial geometrical inequalities can reveal some of the complex properties of higher-dimensional entanglement in multi-partite systems. We provide numerous illustrative applications of this approach, and in particular to a random sample of a thousand density matrices. 
\end{abstract}


\section{Introduction}
\label{intro}

Entanglement is considered  the most non-classical manifestation of quantum mechanics with many strange features such as that the knowledge of the whole system does not include the best possible knowledge of its parts, or it contains correlations that are incompatible with assumptions of classical theories of physics \cite{Horodecki,Horodecki1,Kurzynski}. These qualities resulted in the famous EPR paper and the idea of an alternative theory which was later ruled out by Bell's inequality, and its experimental confirmations \cite{EPR,Bell,Aspect,Hensen,Giustina,Shalm}. Studies of entanglement can be separated into two main categories, efforts regarding the applications of entanglement in quantum protocols and efforts concerning the fundamental questions about the nature of entanglement \cite{Nielsen,Wilde}. 

On the applications side, it was shown that although entanglement in itself does not carry information, it can be helpful in many tasks such as the reduction of classical communication complexity \cite{Cleve}, quantum key distribution \cite{Ekert}, and quantum teleportation \cite{Bennet}. In other words, for one to perform fundamental quantum protocols, entanglement in the form of a maximally entangled state must be consumed \cite{Wilde1}.

In real-world applications, entanglement ordinarily does not come in its pure form but rather as a mixture of pure states; therefore, having a scalable way to detect and quantify entanglement can be important for quantum information processing. This was the motivation for the creation of a class of functions for quantifying entanglement known as entanglement measures, such as quantum discord \cite{zurek}, concurrence \cite{WTT}, squashed entanglement \cite{Matthias}, operators or inequalities for detection of entanglement called entanglement wittiness such as CHSH inequality \cite{CHS}, and partial-transpose criteria \cite{pres,pres1}. All these innovative methods are geared to, and work best for bipartite systems.  Furthermore, despite some efforts to generalizes these methods to multipartite systems \cite{Yang,Mintert,Sarandy}, we are not aware of any efficient way to quantify and detect entanglement in high-dimensional multipartite quantum systems.  We propose one small step in this direction in this manuscript.

As a guide to our construction we use traditional motivations used for most approaches. In particular, it is well accepted that any such measure must satisfy at least a set of three properties: (1) the monotonicity axiom \cite{DiVincenzo}; (2) be vanishing for separable states \cite{Vidal}; and (3) be invariant under local unitary operators.  It is interesting to note that, from the measures and witnesses that we have mentioned above, only concurrence satisfies all of these properties for bipartite systems. 

Entanglement is applicable to more than just quantum information processing. In fundamental physics of entanglement the questions are far more diverse and range from the implications of quantum entanglement and its relations to other parts of physics such as general relativity \cite{Qi,Raamsdonk} to deep philosophical questions related to causality in entangled systems \cite{Nager} and questions regarding mathematical structure of entanglement.  This latter issue is nicely captured in a quote by Bogdan Mielnik ``What picture does one see, looking at a physical theory from a distance, so that the details disappear? Since quantum mechanics is a statistical theory, the most universal picture which remains after the details are forgotten is that of a convex set."\cite{Zyczkowski} We take motivation from this observation in our current work where one particular interpretation of this last questions would be the possibility of the complexity of entanglement arising simply because we are not looking at it using the correct geometrical structures. We assert at a well-motivated  geometry may reduce the properties of entanglement to a set of trivial geometrical properties of convex sets.

As one can guess, answering the last question is of greater importance for the foundation of quantum mechanics and our understanding of physical laws. Still, it can also be beneficial to the problems related to the application of quantum entanglement. If one can simplify the entanglement and eliminate its complexities, he will also be able to quantify and detect it. This is the motivation behind our current manuscript.    

%In this paper, we approach the problem of witnessing and measuring entanglement in large quantum systems from a geometrical point of view. 
%This paper do not aim to give the final answer to either of questions, in other word, we are not claiming we have found a final solution to the problem of quantifying entanglement in large systems, but simply show by introducing geometrical features for quantum systems. 

In this paper, we try to simplify the problem of entanglement by introducing entanglement monotones that play the role of information distances, areas, volumes, and higher-dimensional volumes.\cite{Aslmarand,Miller} This approach enables us to distinguish separable states from entangled states by examining the geometrical differences between them. It also gives us the ability to differentiate between different types of entanglement in quantum systems. However, maybe the most interesting utility of our method would be the ability to `filter' or coarse grain the entanglement in large systems by using inequalities related to higher dimensional geometrical structures without requiring one to calculate all pairwise entanglement of nodes to determine if a specific group of nodes in the system is entangled with the rest of the network. As an observation, we will also show that these geometrical structures will enable us to describe the specific case of monogamy of entanglement as a simple geometric inequality. We do not claim that that the geometric relations that we have defined here can completely characterize the geometry and complexity of quantum information; nevertheless, we do show even  simple geometric constructs can describe and quantify entanglement with more utility than most of the entanglement measures that are currently in use. This can be interpreted as further evidence that by constructing a well-defined geometry, one should be able to reduce the complexity of the problem. 
 
In Sec.~\ref{section: Metric} we will introduce our physical motivation for a new metric called the convoluted metric. Our definition is inspired by the works of Schumacher \cite{Schumacher} and Rolkin and Rajski \cite{Rolkin,Rajski}. We then prove that this metric is a valid distance measure, satisfies all the requisite properties of an entanglement monotone, and the distances $\mathcal{M}_{ij}$ created using this method are indicators of separability between nodes $i$, $j$ and the rest of the system. Later we show, for a tripartite system with at least one separable part, these distances are equivalent to squashed entanglement{\cite{Yang}}. In Sec.~ \ref{section: Volumes}, we will generalize these distances to areas and volumes, as well as higher-dimensional volumes using an approach introduced in \cite{Aslmarand,Miller}, and we show that the areas are also invariant under unitary transformations and is monotonically non-increasing under local operations and classical communication ($LOCC$), and are convex. In Sec.~\ref{APP}, we will show how this approach will offer a new way to detect entanglement beyond the bipartite definition of entanglement and apply it to some relevant  applications. We conclude by suggesting a new function that might be useful for approximating entanglement content of quantum systems. The proof of each proposition will be provided in the appendix. 

\section{Convoluted metric an entanglement monotone}
\label{section: Metric}
Rolkin\cite{Rolkin} and Rajski\cite{Rajski} introduced an information metric 
\begin{equation}
d_{12} = \overline{A_1A_2} = H(\rho_{1|2})+H(\rho_{2|1})
\end{equation} 
between two random variables $A_1$ and $A_2$ with conditional probability density $\rho_{1|2}$, where $H(\rho_{1|2})$ is the usual  conditional entropy. Using this metric, Schumacher \cite{Schumacher} was able to show that geometry created by entangled states has unique geometrical features such as the shortest distance between points in this geometry might not be the direct distance, this has been experimentally shown by \cite{Rezaei} using the measurements on polarization's of entangled photons. Inspired by his work, we introduce two different forms of distance for a random n-partite quantum network with quantum density matrix $\rho_{123...n}$. Hilbert spaces are labeled $A$, $B$, $...$, von Neumann entropy as $S$. So for $\rho_{AB}$, $S(AB)$ is the entropy of the state and $S(A)$ is the entropy of $\Tr_{B}(\rho_{AB})$. Then for $\rho_{ABC}$, the entropy of register $A$ conditioned on register $B$, referred to as the conditional entropy, is:
$$ S(A|B) = S(AB) - S(B) \ . $$
Using this we we define two types of information distance, for $\rho_{ABC}$, the distance, $D_{AB}$ is defined as :
\begin{align}
    D_{AB} = & S(A|B) + S(B|A)  \\
           = & S(AB) - S(B) + [S(AB) - S(A)] \nonumber \\
           = & 2 S(AB) - S(A) - S(B) \nonumber  ,
\end{align}
and the distance $\tilde{D}_{AB}$ is defined as follows:
\begin{align}
    \widetilde{D}_{AB} \equiv & S(A|BC) + S(B|AC)  \\
           = & S(ABC) - S(BC) + [S(ABC) - S(AC)] \nonumber \\
           = & 2 S(ABC) - S(AC) - S(BC) \nonumber  .
\end{align}

\begin{figure}[h!]
\centering
\includegraphics[width=0.3\textwidth]{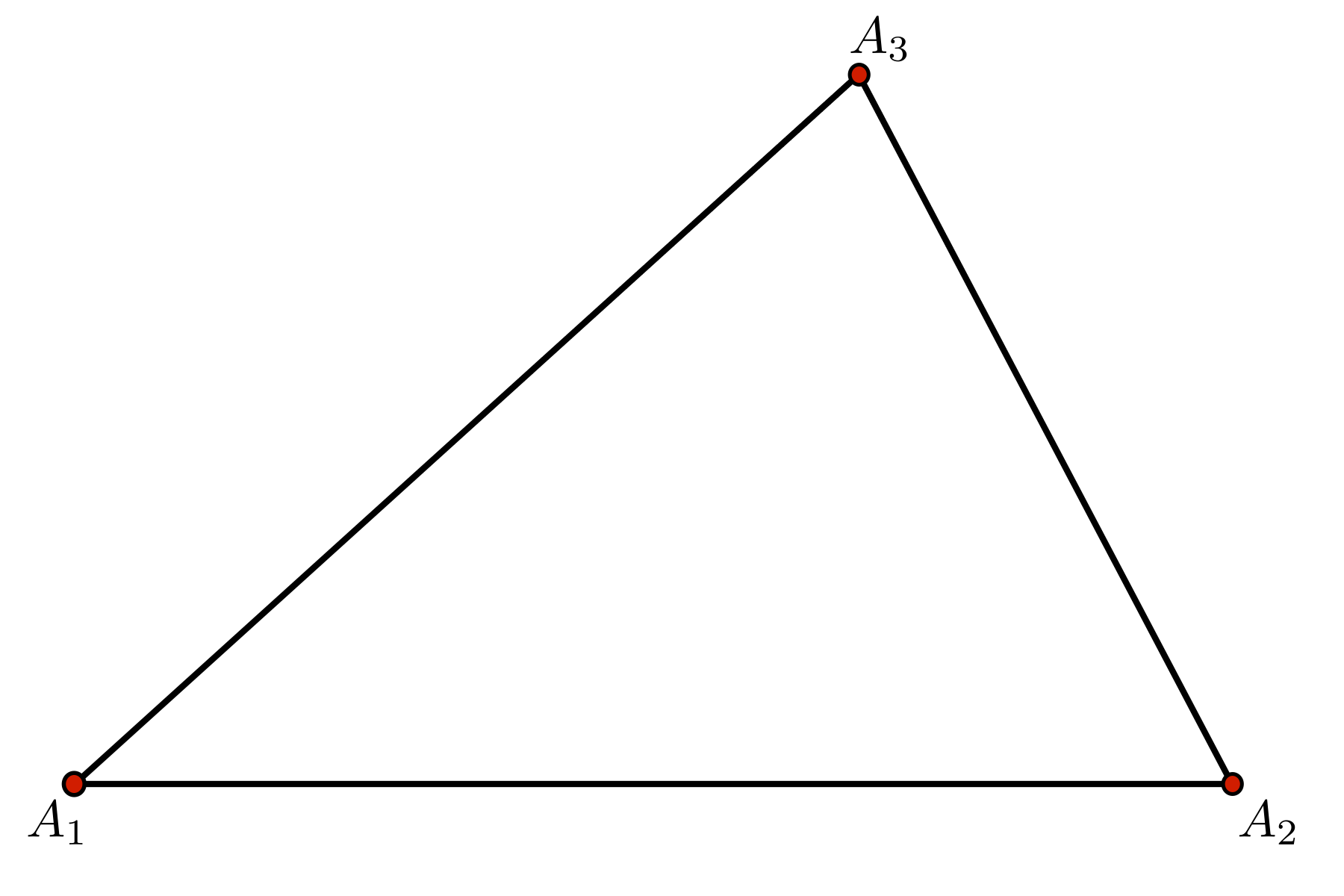}
\caption{Tripartite triangle formed by the three qubits $A_1$, $A_2$ and $A_3$.}\label{fig:triangle}
\end{figure}

The physical motivation for defining such variables is quite simple, lets assume we have a triangle $\overline{A_1A_2A_3}$ illustrated in Fig.~\ref{fig:triangle}.

The distance of vertex $A_1$ and $A_2$ is usually a function of position of $A_1$ and $A_2$ and is independent of position of other points in our geometry. However, if we assume in our geometry this distance will also depend on the position of other vertexes then we must take this into account in our definition of distance. We call this the convoluted distance. This is a non-local effect in the sense it is a measure of the asymmetry between $A_1$ and $A_2$ with respect to extra resource of the register $A_3$.

Consequently, we define our metric $M_{AB}$ (Convoluted metric) for $\rho_{ABC}$ as difference of the distances $D$ and $\tilde D$:
\begin{align}
\label{eq:M}
    M_{AB} \equiv & D_{AB} - \widetilde{D}_{AB}  \\
           = & 2S(AB) - S(A) -S(B) - [2S(ABC) - S(AC) - S(BC)] \nonumber \\
           = & -2S(ABC) + 2S(AB) + S(AC) + S(BC) -S(A) - S(B) \nonumber  .
\end{align}

As one might think for the definition, this metric $M_{AB}$ provides the means of non-separability of $\rho_{AB}$ from $\rho_C$, since it measures distance of Alice and Bob's registers both locally and non-locally.

%\begin{proposition}
%\label{thm:metric}
For any density matrix $\rho_{ABC}$, the convoluted metric $M_{AB}$ may be seen as a pseudo-metric. That is to say:
\begin{enumerate}
   \item $M_{AB} = M_{BA} $,
  \item $M_{AB} \geq 0\ \hbox{and is equal to}\ 0\ $\hbox{iff}\ $ \rho_{ABC}=\rho_{AB} \otimes \rho_C $, ~and
   \item $M_{AB} + M_{BC} \geq M_{AC} $.
   \end{enumerate}
%\end{proposition}

For the special case of tripartite density matrix in form of $\rho_{ABC}=\rho_{AB} \otimes \rho_C $,  $M_{AB}$ is equal to the definition of bipartite squashed entanglement \cite{Yang}.  In other word one can think of this metric as geometrical representation of squashed entanglement in this case.  

 It's easy to show that convoluted metric $M_{AB}$ is an entanglement monotone  and  satisfies the following properties:
\begin{enumerate}
    \item $M_{AB}$ is invariant under local and global isometries,
    \item $M_{AB}$ is non-increasing under LOCC,
    \item $M_{AB}$ is Convex. 
\end{enumerate}

So far what we have discussed are related to pure states, however; given a state $\rho_{ABX}$, one could generally find many ways of decomposing it into an ensemble $\{p_{i},\rho_{i}\}_{i \in \Sigma}$. In other words, there are many ensembles whose average state is $\rho_{ABX}$. The decomposition may vary both in the number of states in the ensemble, $|\Sigma|$, and the choice of state $\rho_{i}$. This suggests an extra measure must be taken to expand the definition for mixed density matrices. Therefore for Given $\rho_{ABX}$, the convoluted metric, $M_{AB}$, for a mixed quantum density matrix is defined as follows:
\begin{align}
    M_{AB} \equiv \inf \{ \sum_{j} p_{j} M_{AB}(\rho^j_{ABX}) \}
\end{align}
where the infimum is taken over all possible decomposition's of pure state density matrix $\rho^j_{ABX}$.

Now that we have proved that our metric can be a useful tool in investigating separability problem in quantum systems, we proceed by developing new geometrical features based on this metric.

\section{Higher-dimensional structures}
\label{section: Volumes}
In order to study and characterize higher-dimensional entanglement networks, we are interested in coarse-graining quantum networks.  One example of questions one needs to answer for coarse-graining is if we have a density matrix of the form $\rho_{ABCD}= \rho_A \otimes \rho_{BCD}$ using the distances and squashed entanglement we can only state that  $\rho_{ABCD}$ is non separable or entangled but the type of entanglement such as bipartite or tripartite needs extra calculations. For such purposes we introduce higher dimensional structures such as areas and volumes. 

For $\rho_{A_1...A_n}$ following the procedure for distances, we define two different areas , the $Area_{A_iA_jA_k}$ and $\widetilde{Area}_{A_iA_jA_k}$. Which are defined as followed:
\begin{align}
Area_{A_iA_jA_k}= -\big{[}&S(\rho_{A_i|A_jA_k})*S(\rho_{A_j|A_iA_k})\\
+&S(\rho_{A_i|A_jA_k})*S(\rho_{A_k|A_jA_i}) \nonumber\\
+&S(\rho_{A_k|A_iA_j})*S(\rho_{A_j|A_jA_k})\big{]} \nonumber
\end{align} 
\begin{align}
    \widetilde{Area}_{A_iA_jA_k} = -\big{[}&S(\rho_{A_i|A_jA_k..A_n})*S(\rho_{A_j|A_iA_k...A_n}) \\
    +&S(\rho_{A_i|A_jA_k...A_n})*S(\rho_{A_k|A_jA_i...A_n}) \nonumber\\
    +&S(\rho_{A_k|A_iA_j...A_n})*S(\rho_{A_j|A_jA_k....A_n})\big{]} \nonumber
\end{align}

Then using this two different areas, we define our convoluted area $^2M_{A_iA_jA_k}$ for $\rho_{A_1...A_n}$  as:
\begin{equation}
^2M_{A_iA_jA_k}=Area-\widetilde{Area},
\end{equation} 

and later we expand this to mixed states by taking the infimum over all possible decompositions of density matrix. Therefore for a mixed density matrix  $\rho_{A_1...A_n}$, the convoluted area $^2M_{A_iA_jA_k}$ will equal to
\begin{equation}
^2M_{A_iA_jA_k}=\inf \left [ \sum_j p_j ~~{}^2{M}_{A_iA_jA_k}(\rho^j_{A_1...A_n})\right] \geq 0
\end{equation} 
We conjecture that for $\rho_{A_1...A_n}$ the convoluted area ${}^2M_{A_iA_jA_k}$ is convex.  Furthermore, in Appendix $C$ we proved that this area $^2M_{A_iA_jA_k}$ satisfies the following three properties: (1) it is invariant under local unitary operators; (2) it will vanish if subsystems $A_iA_jA_k $ are separable from the rest of the system; and (3)  it is non-increasing under LOCC.

We can naturally generalize this to volumes and higher-dimensional volumes using definition introduced recently in \cite{Miller}. Consequently, given $\rho_{A_1...A_n}$ two type of volumes  $^{(m-1)}V_{A_1A_2\ldots A_m}$  and $^{(m-1)}\widetilde{V}_{A_1A_2\ldots A_m}$  are defined as:
\begin{equation}
\label{eq:nV}
\begin{array}{ll}
{}^{^{(m-1)}}\!{\mathcal V}_{A_1A_2\ldots A_m}& := (-1)^m
\displaystyle
\sum_{a_1,a_2,\ldots a_m=A_1}^{A_m} 
\left(  \frac{1+\epsilon_{a_1a_2\ldots a_m}}{2}  \right)
\underbrace{
S_{a_1|a_2\ldots a_{m}}S_{a_2|a_1a_3\ldots a_{m}} \ldots S_{a_{m-1}|a_1a_2\ldots a_{m\!-\!} a_{m}}
}_{
\hbox{\tiny product of}\ m\ \hbox{\tiny conditional entropies}}, 
\end{array}
\end{equation}
and 
\begin{equation}
\label{eq:nV}
\begin{array}{ll}
{}^{^{(m-1)}}\!\widetilde{{\mathcal V}}_{A_1A_2\ldots A_m}& := (-1)^m
\displaystyle
\sum_{a_1,a_2,\ldots a_m=A_1}^{A_m} 
\left(  \frac{1+\epsilon_{a_1a_2\ldots a_m}}{2}  \right)
\underbrace{
S_{a_1|a_2\ldots a_{n}}S_{a_2|a_1a_3\ldots a_{n}} \ldots S_{a_{m-1}|a_1a_2\ldots a_{m\!-\!} a_{n}}
}_{
\hbox{\tiny product of}\ m\ \hbox{\tiny conditional entropies}}.
\end{array}
\end{equation}
This allows us to define, the $m$-dimensional convoluted volume $^{(m-1)}M_{A_1A_2\ldots A_m}$ as:
\begin{equation}
^{(m-1)}{\mathcal M}_{A_1A_2\ldots A_m} :=  {{}^{^{(m-1)}}\!{\mathcal V}_{A_1A_2\ldots A_m}-{}^{^{(m-1)}}\!{\widetilde{\mathcal V}}_{A_1A_2\ldots A_m}}
\end{equation}
and again we can expand this to mixed density matrices as:  
\begin{equation}
^{(m-1)}{\mathcal M}_{A_1A_2\ldots A_m} :=  \inf\left( \sum_j p_j ~^{(m-1)}{\mathcal M} (\rho_j)\right)
\end{equation}
where the infimum is taken over all possible decompositions.

Again, as conjecture we suggest that  the  ${}^mM_{A_iA_jA_k..A_m}$  is convex. Its also possible for these higher dimensions to show that this convoluted volume is (1) invariant under local and global isometries, and (2) is non-increasing under LOCC.

Now that we showed that these structures are entanglement monotones, in the next section we will examine a few applications of these entanglement monotones, and highlight their potential utility. 

\section{Illustrative applications}
\label{APP}

\subsection{Filtering the entanglement in quantum networks}
Filtering the entanglement is of significant importance for quantum computing purposes in large quantum networks. One of the applications of our convoluted structures is the ability to Filtering the entanglement. It's possible in future that there will be services offering cloud quantum computing. Let's assume, we have a quantum cloud system of N nods, which we call $\Omega_N$. Since, these types of technologies will be used by multiple users, its necessary to be able to filter entanglement to avoid the disruption and leakage of information from nodes being used by one user to the other. Therefore, it's essential to be able to find islands of entangled nodes that are separable from rest of system. One can rephrase this question in this way " is it possible to find out if a group of $m$ nods $\Omega_m \subseteq \Omega_N$ is entangled to the rest of systems without calculating pair-wise entanglement''?. One can answer this by using the $m-1$ dimensional convoluted volumes. This can be expressed as the following observation.
 
 \begin{observation}
For the a density matrices $\rho_{A_1...A_N}$ to find if a group of nods ${A_i|i\in L} $ is entangled to rest of system one have to calculate the ${}^{|L|-1}M_{A_i,i \in L}$, if the systems is of form $\rho_{A_1...A_N}=\rho_{A_i,i \in L} \otimes \rho_{A_j,j\notin L}$ then  ${}^{|L|-1}M_{A_i,i \in L}$ will vanish. Therefore by using these higher dimensional structures  one can filter the entanglement in quantum network without a need to calculate all the bipartite entanglements. 
\end{observation}

\subsection{Categorizing the entanglement}  
Second illustrative application that we want to present, is the ability of these entanglement monotones to differentiate between different types of entanglement. Let's imagine we have a quantum state $\rho_{ABCD}$.  
While the joint quantum state of $ABCD$ that factors into a
product, one for A, B, C, and D is fully
separable (and so are mixtures of these products), one can have a two bi-partite non-separable state as $\rho_{AB} \otimes \rho_{CD}$ or tripartite non-separable state, etc. We can 
question, regardless of the entanglement content of these quantum systems; how can we find out what type of entanglement is present? We show here that by using convoluted structures, we can answer this question. 

\begin{observation}
For the two density matrices $\rho_{ABCD}=\rho_{AB} \otimes \rho_{CD}$ and $\tilde\rho_{ABCD}=\rho_{ABC} \otimes \rho_{D}$ with the same entanglement content one can easily differentiate between the type of entanglement by looking at their different geometrical structure in terms of areas, 
\begin{align}
&^{2}M_{ABC}(\rho)\neq 0 ~~~~M_{AB}(\rho)=M_{CD}(\rho)=0,\\
&^{2}M_{ABC}(\tilde\rho)= 0 ~~ ~~M_{AB}\neq 0,  M_{CD}\neq0. \nonumber
\end{align}
This illustrates that the type of entanglement in the $\rho$ is bipartite entanglement and  $\tilde\rho$ is of tripartite nature.
\end{observation}

\subsection{Simplifying complex properties of entanglement}  
The third application of these structures can be the ability to simplify the complex properties of entanglement to a set of trivial geometrical features. As an example, here we show that special cases of entanglement monogamy can be reduced to a well-known trivial geometrical inequality known as Ono's therm  \cite{Ono}. 
Ono's Theorem states, that for  a triangle ABC with acute or right angles we have 
 
\begin{align}
27\left[a^2+b^2-c^2\right]^2\left[a^2+c^2-b^2\right]^2\left[c^2+b^2-a^2\right]^2 \leq (4A)^6
\end{align}

\begin{observation}
For a state $\rho_{ABC}$ If two qubits A and B are maximally correlated they cannot be correlated at all with a third qubit C.

{\bf{Proof}}. We will assume that for a $\rho_{ABCD}=\rho_{ABC} \otimes \rho_D$, parts $A$, $B$ are maximally entangled to each other and they are also to some degree entangled to C meaning 
\begin{equation}
    M_{AB} \neq 0
\end{equation}
 we will show this will lead to violation of Ono's inequality, therefore two maximally entangled qubits can't be correlated at all with a third qubit. We know that for $\rho_{ABCD}$  our convoluted Area
\begin{equation}
    ^2M_{ABC}=0 
\end{equation}
since $\rho_{ABC}$ is separable from $\rho_D$. Therefore 

\begin{align}
\left[M_{AB}^2+M_{AC}^2-M_{BC}^2\right]^2\left[M_{AB}^2+M_{BC}^2-M_{AC}^2\right]^2\left[M_{AC}^2+M_{BC}^2-M_{AB}^2\right]^2 \leq 0
\end{align}

 so one or all of the terms in the right hand side of the inequality must be zero
 \begin{align}
 &M_{AB}^2+ M_{AC}^2- M_{BC}^2 =0 \\ 
 &M_{BC}^2+ M_{AB}^2- M_{AC}^2 =0 \\
 &M_{BC}^2+ M_{AC}^2- M_{AB}^2 =0 
 \end{align}
Now using the fact that since $\rho_{AB}$ is maximally entangled and symmetric then 
\begin{align}
    M_{AC}=M_{BC} \neq 0 
\end{align}
 This will reduce the equations to
 \begin{align}
     &M_{AB}=0 \\
     & or \nonumber\\
     &M_{AB}^2= 2M_{AC}^2
 \end{align}
but the second equation is not possible since $M_{AC}$ is the maximum value that M can take since we already assumed that $A $ and $B$ are maximally entangled, therefore $M_{AB}$ must be zero which is contradiction and this proves our claim. 
\end{observation}

\subsection{Approximating entanglement content of quantum systems}  
As the fourth and final use case, we will try to harvest these geometrical structures to approximate the entanglement content of the quantum system. The argument is that the entanglement content of a tripartite density matrix is the sum of bipartite entanglement of quantum systems and the entanglement shared between three parts, using this logic for the n-partite system, we can suggest the following. 
Given $\rho_{A_1...A_n}$, the entanglement content of system can be approximated by $E$:
\begin{equation}
    E(\rho_{a_1a_2..a_m})=\frac{1}{{2!}}\sum_{a_ia_j}M_{a_ia_j}+\frac{1}{{3!}}\sum_{a_ia_ja_k}{}^3M_{a_ia_ja_k}+\frac{1}{{4! }}\sum_{a_ia_ja_ka_l}{}^4M_{a_ia_ja_ka_l}+...
\end{equation}
The coefficient are normalization factors to avoid the multiple counting. For mixed quantum density matrix 
\begin{equation}
    \rho_{a_1a_2..a_m} =\sum_i \lambda_i \rho^i_{a_1a_2..a_m}
\end{equation}
will equal to 
\begin{equation}
    E(\rho_{a_1a_2..a_m})= inf\big{(}\sum_n \lambda_i E(\rho^i)\big{)}.
\end{equation}
$E(\rho)$  satisfies the following two properties: (1) it is invariant under local and global isometries; and (2) it is non-increasing under LOCC.  Furthermore,  $E(\rho)$ is also convex due to convexity of each of it's individual parts. By way of illustration of this point, we can demonstrate that the $E(\rho)$ function canapproximate the entanglement content of a quantum system.  For example, we analyzed the entanglement content of quantum networks with four parts. Because we want to compare our hypothesis to concurrence $(C)$,  we generated $1000$ random density matrices of form $\rho_{12} \otimes \rho_{34}$ using \cite{QETLAB}.   The entanglement content of this system would be equal to $C(\rho_{12})+ C(\rho_{34})$. Next, we calculate the values of $E(\rho_{1234})$ and normalize them by putting $E_{\rho_{bell}\otimes \rho_{bell}}$ equal to 2 ( we divide all the values by $\frac{E_{\rho_{bell}\otimes \rho_{bell}}}{2}$ ) . After this we sorted the density matrices based on the value of concurrence and plot the values of concurrence and $E$.  Fig.~2 illustrates that $E$ can be used to approximate the entanglement content of quantum systems. 
\begin{figure}[h]
\label{figl}
\centering
\includegraphics[width=0.9\textwidth]{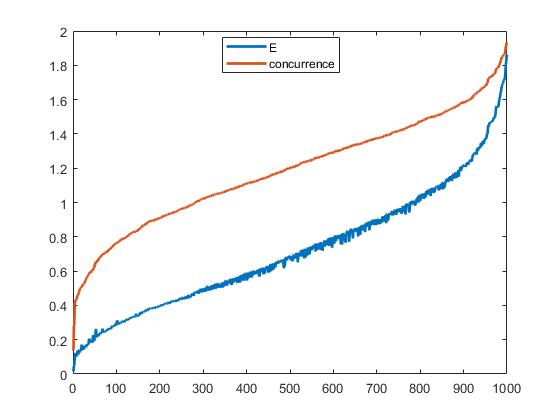}
\caption{Comparison of concurrence and $E(\rho_{1234})$ for a random sample of a thousand density matrices of the form $ \rho_{1234}= \rho_{12} \otimes \rho_{34}$. We normalized $E$ by setting  $E_{\rho_{bell}\otimes \rho_{bell}} =2$ .}\label{fig:graph}
\end{figure}

We conjecture that replacing the Von Neumann entropy with Shannon entropy in the definitions of metric and volumes, one will be able to generate an experimental lower bond for the entanglement that would be useful for applications such as quantum optics and information theory.

\section{Conclusion}

In this work, we attempted to simplify the problem of multipartite entanglement by introducing entanglement monotones that play the role of distances, areas, and higher-dimensional information volumes. This approach enables us to distinguish separable states from entangled states by examining their geometrical differences through inequalities --- a sort of `filtering of quantum entanglement.'  To us, the most interesting utility of our method is to filter entanglement without requiring an exponentially increasing number of calculations to determine if a specific group of nodes in the system is entangled with the rest of the network. As an observation, we also showed that these geometrical structures will enable us to describe the specific case of monogamy of entanglement as a simple geometric inequality. We do not claim that that the geometry we have created is a complete geometry of quantum information; nevertheless, we show even such a simple geometric constructs can describe and quantify entanglement with more utility than most of the entanglement measures that are currently in use. This can be interpreted as further evidence that by constructing a well-defined geometry, one should be able to reduce the complexity of the higher dimensional entangled systems. We suggest by exploring these higher-dimensional entropic geometries one can gain a deeper insight into some of the strange properties of entangled networks, though realizing that this approach offers but another independent glimpse into entanglement.

\section{Acknowledgment}

SMA would like to thank Ian George for his suggestions and helpful discussions. WAM would like thank support from the Air Force Office of Scientific Research, Air Force Research Laboratory's Information Directorate as well as L3Harris. This research was supported in part by the Air Force Research Laboratory Information Directorate , through the Air Force Office of Scientific Research Summer Faculty Fellowship Program®, Contract Numbers $\#FA8750-15-3-6003$, $\#FA9550-15-0001$ and $\#FA9550-20-F-0005$, AFOSR/AOARD grant $\#FA23861714070$, AFOSR/DURIP grant $\#FA95501910389$ and  support from L3Harris. PMA for this work would like thank support from the Air Force Office of Scientific Research.  Any opinions, findings, conclusions or recommendations expressed in this material are those of the author(s) and do not necessarily reflect the views of AFRL.

\section{Appendix}
In this section we will present the proves for our proposed claims in the manuscript.

\subsection{Appendix A}
For any density matrix $\rho_{ABC}$, the convoluted metric $M_{AB}$ may be seen as a pseudo-metric. That is to say:
\begin{enumerate}[{1.}1]
   \item $M_{AB} = M_{BA} $
  \item $M_{AB} \geq 0\ \hbox{and is equal to}\ 0\ $\hbox{iff}\ $ \rho_{ABC}=\rho_{AB} \otimes \rho_C $.
   \item $M_{AB} + M_{BC} \geq M_{AC} $
   \end{enumerate}

{\bf{Proof}}. 
\begin{enumerate}[{1.}1]
\item  In $Eq.4$  A and B are interchangeable. 
\item 
We know form  strong subadditivity of quantum entropy (SSA) inequality  \cite{Lieb} that for tripartite separable density matrix  
\begin{equation}
S(\rho_{ABC})\leq S(\rho_{AC})+S(\rho_{BC})-S(\rho_{C})
\end{equation}
One can write this inequality in two different ways
\begin{equation}
S(\rho_{ABC})\leq S(\rho_{AC})+S(\rho_{AB})-S(\rho_{A})
\end{equation}
\begin{equation}
S(\rho_{ABC})\leq S(\rho_{BC})+S(\rho_{AB})-S(\rho_{B})
\end{equation}

Now by summing up these inequalities one will reach to 

\begin{equation}
S(\rho_{ABC})+ S(\rho_{ABC})\leq S(\rho_{BC})+S(\rho_{AB})-S(\rho_{B})+S(\rho_{AC})+S(\rho_{AB})-S(\rho_{A})
\end{equation}
This will lead to 

\begin{equation}
\big{[}S(\rho_{ABC})-S(\rho_{BC})\big{]}+ \big{[}S(\rho_{ABC})-S(\rho_{AC})\big{]} \leq S(\rho_{AB})-S(\rho_{B})+S(\rho_{AB})-S(\rho_{A})
\end{equation}

Which is equal to 
\begin{equation}
\widetilde{D}_{AB} \leq D_{AB},
\end{equation}
hence $M_{AB}$ is  
\begin{equation}
0\leq D_{AB}-\widetilde{D}_{AB}= M_{AB}.
\end{equation}
 further more, $M_{AB}$ is zero if and only if $ \rho_{ABC}=\rho_{AB} \otimes \rho_C  $. Since 
 \begin{align}
 \label{eqSep}
 S(ABC)=S(AB)+S(C) \\
 S(AC)=S(A)+S(C) \nonumber\\
 S(BC)=S(B)+S(C) \nonumber
 \end{align}
 for $ \rho_{ABC}=\rho_{AB} \otimes \rho_C  $ and plugging in \ref{eqSep} in $Eq.~4$ will make $M_{AB}=0$.
\item For triangle inequality we start with definition of $ M_{AB}$ 

\begin{align}
&M_{AB}= S(\rho_{A|B})+S(\rho_{B|A})-S(\rho_{A|BC})-S(\rho_{B|AC})\\
&M_{AB}=2S(\rho_{AB})-2S(\rho_{ABC})+S(\rho_{BC})+S(\rho_{AC})-S(\rho_A)-S(\rho_B)
\end{align} 

Now plug-in in the definition into 
\begin{equation}
M_{AB}+M_{BC}\geq M_{AC}
\end{equation}
 and after canceling the terms we have
\begin{equation}
2S(\rho_{AB})-2S(\rho_{ABC})+2S(\rho_{BC})-2S(\rho_B) \geq 0 .
\end{equation}
This is strong subadditivity equation \cite{Lieb}
\begin{equation}
    S(\rho_{AB})+S(\rho_{BC}) \geq S(\rho_{ABC})+S(\rho_B) 
\end{equation}
and is true for any arbitrary density matrix.$\square $

 \end{enumerate}

For n-partite systems to show $M_{12}$ is a metric, one just need to adjust $\rho_{3}$ to $\rho_{3...n}$.

\subsection{Appendix B}

Given $\rho_{ABX}$, the convoluted metric $M_{AB}$ is an entanglement monotone for detecting separability between the registers $AB$ and $X$  satisfies the following properties:
\begin{enumerate}[{2.}1]
    \item $M_{AB}$ is invariant under local and global isometries
    \item $M_{AB}$ is non-increasing under LOCC
    \item $M_{AB}$ is Convex. 
\end{enumerate}

{\bf{Proof}}. If we write the $M_{AB}$  in term of conditional Mutual information 

\begin{align}
    M_{AB} = & -2S(ABC) + 2S(AB) + S(AC) + S(BC) -S(A) - S(B) \\
           = & [S(AB) + S(AC) -S(ABC) - S(A)] + [S(AB) + S(BC) -S(ABC) - S(B)] \nonumber \\
           = & I(B:C|A) + I(A:C|B) \nonumber \label{defn:MasCMI}
\end{align}

\begin{enumerate}[{2.}1]
\item Von Neumann entropy is invariant under unitary operation therefore $M_{AB}$ is invariant under unitary operation. 
\item Conditional mutual information is non-increasing under $LOCC$  \cite{Yang} therefore $M_{AB}$ is non-increasing under $ LOCC$. 
\item Conditional mutual information is convex \cite{Yang} therefore, $M_{AB}$ is convex since the sum of two convex functions is also convex. 
\end{enumerate}
$\square$
\subsection{Appendix C}
\label{APC}

Given $\rho_{A_1...A_n}$, the  $^2M_{A_iA_jA_k}$  satisfies the following properties:
\begin{enumerate}[{3.}1]
    \item  is invariant under local and global isometries
    \item  is Convex.
     \item  is non-increasing under LOCC.
\end{enumerate}

{\bf{Proof}}.
\begin{enumerate}[{3.}1]
\item The von Neumann entropy is invariant under unitary operation therefore $^2M_{A_i A-j A_k}$ is invariant under unitary operation. 
\item
By adding and subtracting the terms 
\begin{align}
& S(\rho_{A_1|A_2A_3})*S(\rho_{A_2|A_1A_3A_{4 ....n}})\\
 &S(\rho_{A_2|A_1A_3})*S(\rho_{A_3|A_1A_2A_{4 ....n}})\\
 &S(\rho_{A_3|A_1A_2})*S(\rho_{A_1|A_2A_3A_{4 ....n}})
\end{align} 
To $^2M_{A_1A_2A_3}$ make it equal to 
\begin{align}
^2M_{A_1A_2A_3}=-\bigg{\{}& S(\rho_{A_1|A_2A_3})*\big{[}S(\rho_{A_2|A_1A_3})-S(\rho_{A_2|A_1A_3A_{4 ....n}})\big{]}\\
+&S(\rho_{A_2|A_1A_3A_{4 ....n}})*\big{[}S(\rho_{A_1|A_2A_3})-S(\rho_{A_1|A_2A_3A_{4 ....n}})\big{]} \nonumber\\
+& S(\rho_{A_2|A_1A_3})*\big{[}S(\rho_{A_3|A_1A_2})-S(\rho_{A_3|A_1A_2A_{4 ....n}})\big{]} \nonumber\\
+&S(\rho_{A_3|A_1A_2A_{4 ....n}})*\big{[}S(\rho_{A_2|A_1A_3})-S(\rho_{A_2|A_1A_3A_{4 ....n}})\big{]} \nonumber\\
+& S(\rho_{A_3|A_1A_2})*\big{[}S(\rho_{A_1|A_2A_3})-S(\rho_{A_1|A_2A_3A_{4 ....n}})\big{]} \nonumber\\
+&S(\rho_{A_1|A_2A_3A_{4 ....n}})*\big{[}S(\rho_{A_3|A_1A_2})-S(\rho_{A_3|A_1A_2A_{4 ....n}})\big{]}\bigg{\}} \nonumber
\end{align} 

Then we rewrite this as 

\begin{align}
^2M_{A_1A_2A_3}= -&\big{[}S(\rho_{A_1|A_2A_3})+S(\rho_{A_3|A_1A_2A_{4 ....n}})\big{]}*\big{[}S(\rho_{A_2|A_1A_3})-S(\rho_{A_2|A_1A_3A_{4 ....n}})\big{]}\\
-& \big{[}S(\rho_{A_3|A_1A_2})+S(\rho_{A_2|A_1A_3A_{4 ....n}})\big{]}*\big{[}S(\rho_{A_1|A_2A_3})-S(\rho_{A_1|A_2A_3A_{4 ....n}})\big{]} \nonumber\\
-& \big{[}S(\rho_{A_2|A_1A_3})+S(\rho_{A_1|A_2A_3A_{4 ....n}})\big{]}*\big{[}S(\rho_{A_3|A_1A_2})-S(\rho_{A_3|A_1A_2A_{4 ....n}})\big{]} \nonumber
\end{align} 
One can express this in term of Conditional Mutual Information as 
\begin{align}
^2M_{A_1A_2A_3}=& \big{[}I(A_1;A_2A_3)+I(A_3;A_1A_2A_{4 ....n})-S(A_1)-S(A_3)\big{]}*\big{[}I(A_2;A_{4 ....n}|A_1A_3)\big{]}\\
+& \big{[}I(A_3;A_1A_2)+I(A_2;A_1A_3A_{4 ....n})-S(A_3)-S(A_2)\big{]}*\big{[}I(A_1;A_{4 ....n}|A_2A_3)\big{]} \nonumber\\
+& \big{[}I(A_2;A_1A_3)+I(A_1;A_2A_3A_{4 ....n})-S(A_1)-S(A_2)\big{]}*\big{[}I(A_3;A_{4 ....n}|A_1A_3)\big{]} \nonumber
\end{align} 

Now since we know subsystems  $A_iA_jA_k $ is separable from the rest of the system $^2M_{A_iA_jA_k}$ will equal to zero due to $I(A_i;A_{4 ....n}|A_jA_k)=0$.
\item To prove that $^2M_{A_1A_2A_3}$ is non-increasing under $LOCC$  we have to use the proposition by \cite{Yang} 
\begin{proposition}
A convex function $f$ does not increase under $LOCC$ if and only if
\begin{enumerate}
\item f is invariant under local unitary operators
\item if f satisfies 
\begin{equation}
f\left(\sum_i p_i \rho_{AB}^i\otimes |i\rangle x \langle i| \right)=\sum_i p_if(\rho_{AB}^i).
\end{equation}
\end{enumerate}
\end{proposition}
Since $^2M_{A_1A_2A_3}$ satisfies both of these, then it is non-increasing under $LOCC$.
\end{enumerate}

\subsection{Appendix D}
Given $\rho_{A_1...A_n}$, the  $E(\rho)$  satisfies the following properties:
\begin{enumerate}[{6.}1]
    \item   $E(\rho)$ is invariant under local and global isometries;
    \item  $E(\rho)$ is non-increasing under LOCC; and
    \item  $E(\rho)$ is Convex.
\end{enumerate}

{\bf{Proof}}.
\begin{enumerate}
\item $E(\rho)$ is invariant under local unitary operators due to $^mM_{A_i...A_m}$ being invariant under local unitary operators. 
\item From our previous discussions,  we know that $^mM_{A_i...A_m}$ are non-increasing under $LOCC$, and we know that sum of non-increasing terms will also be non-increasing under $LOCC$. Then $E(\rho)$ is non increasing  under $LOCC$.
\end{enumerate}
\end{document}